\pdfoutput=1
\documentclass[12pt]{article}
\usepackage{graphicx,psfrag,epsf}
\usepackage{amsmath}
\usepackage{amssymb}
\usepackage{enumerate}
\usepackage{enumitem}
\usepackage{amsfonts}
\usepackage{amsthm}
\usepackage{natbib}
\usepackage{setspace}
\usepackage{color}
\usepackage{array}
\usepackage{fullpage}
\usepackage{bm}
\usepackage{bbm}
\usepackage{mathtools}
\usepackage[toc,page]{appendix}
\usepackage{float}
\usepackage[usenames,dvipsnames]{xcolor}
\usepackage{subcaption}
\usepackage{authblk}
\usepackage{multirow}
\usepackage{footnote}
\usepackage[hidelinks, colorlinks=true, linkcolor=black, citecolor=blue]{hyperref}

\newtheorem{theorem}{Theorem}
\newtheorem{lemma}{Lemma}

\newtheorem{definition}{Definition}
\newtheorem{remark}{Remark}

\numberwithin{equation}{section}

\allowdisplaybreaks
\begin{document}
	
	\title{Penalized Maximum Tangent Likelihood Estimation and Robust Variable Selection}
	
	\author{Yichen Qin, Shaobo Li, Yang Li, and Yan Yu\thanks{Yichen Qin is Assistant Professor (\href{mailto:qinyn@ucmail.uc.edu}{qinyn@ucmail.uc.edu}), Department of Operations, Business Analytics, and Information Systems, Carl H. Lindner College of Business, University of Cincinnati, Cincinnati, OH 45221; Shaobo Li is PhD candidate (\href{mailto:lis6@mail.uc.edu}{lis6@mail.uc.edu}), Department of Operations, Business Analytics, and Information Systems, Carl H. Lindner College of Business, University of Cincinnati, Cincinnati, OH 45221; Yang Li is Associate Professor (\href{mailto:yang.li@ruc.edu.cn}{yang.li@ruc.edu.cn}), School of Statistics and Center for Applied Statistics, Renmin University of China, Beijing 100872; Yan Yu is Joseph S. Stern Professor of Business Analytics (\href{mailto:yan.yu@uc.edu}{yan.yu@uc.edu}), Department of Operations, Business Analytics, and Information Systems, Carl H. Lindner College of Business, University of Cincinnati, Cincinnati, OH 45221}}
	
	\maketitle

		\begin{abstract}
			
			\setlength{\baselineskip}{0.5cm}
			
			We introduce a new class of mean regression estimators --- 			
			penalized maximum tangent likelihood estimation --- for 
			high-dimensional regression estimation and variable selection.  
			We first explain the motivations for the key 
			ingredient, maximum tangent likelihood estimation (MTE), and 
			establish its asymptotic properties. 
			We further propose a penalized MTE for variable 
			selection and show that it is 
			$\sqrt{n}$-consistent, enjoys the oracle property. 
			The proposed class of estimators consists penalized $\ell_2$ 
			distance, penalized exponential squared 
			loss, penalized least trimmed square and penalized least square as 
			special cases and can be regarded as a mixture of minimum 
			Kullback-Leibler distance estimation and minimum $\ell_2$ distance 
			estimation. 
			Furthermore, we consider the proposed class of estimators under 
			the high-dimensional setting when the number of variables $d$ can 
			grow exponentially with the sample size $n$, and show that the 
			entire class of estimators (including the aforementioned 
			special 
			cases) can achieve the optimal rate of 
			convergence in the order of $\sqrt{\ln(d)/n}$. 
			Finally, simulation studies and real 
			data analysis demonstrate the advantages of the penalized 
			MTE.			
		\end{abstract}


	%%%%%%%%%%%%%%%%%%%%%%%%%%%%%%%%%%%%%%%%%%%%%%%%%%%%%%%%%%%%%%%%%%%%%%%%%%%%%%%%%%%%%
	
	\section{Introduction}\label{sec.intro}
	Selecting explanatory variables has become one of the most important tasks 
	in statistics.  However, many of existing variable selection methods are 
	sensitive to outliers.  To address this issue, we develop a class of robust 
	linear regression estimators, namely, penalized maximum tangent likelihood 
	estimation.
	
	Existing popular variable selection methods include Lasso 
	\citep{Tibshirani1996}, SCAD \citep{Fan2001}, and adaptive-Lasso 
	\citep{Zou2006}.  Their properties in the high-dimensional regression 
	setting are extensively studied 
	\citep{fan2004,meinshausen2006,Bickel2009}.  A unified 
	theoretical framework of penalized high-dimensional methods was provided in 
	\citet{Negahban2012}.  
	Many aforementioned methods can be expressed as penalized likelihood 
	estimation (assuming normal distributions for the random errors),
	$\check{\boldsymbol{\beta}}=\arg\max_{\boldsymbol{\beta}} \left\{ 
	\sum_{i=1}^{n}\ln f({\bf z}_i;\boldsymbol{\beta}) - n\sum_{j=1}^{d} 
	p_{\lambda}(\beta_j) \right\}$ where $\{{\bf z}_i\}_{i=1}^n=\{y_i,{\bf 
	x}^T_i\}_{i=1}^n$ represents the response variable and covariates, and $f$ 
	represents the normal distribution with zero mean, and $f({\bf 
	z}_i;\boldsymbol{\beta})=f(y_i-{\bf x}_i^T 
	\boldsymbol{\beta})$ (note we use $f({\bf z}_i;\boldsymbol{\beta})$ and 
	$f(y_i-{\bf x}_i^T \boldsymbol{\beta})$ interchangeably in this article).  
	However, the performance of such an estimator usually 
	degrades drastically even if a small proportion of data is contaminated.
	
\begin{figure}
		\centering 		
		\includegraphics[scale=0.5]{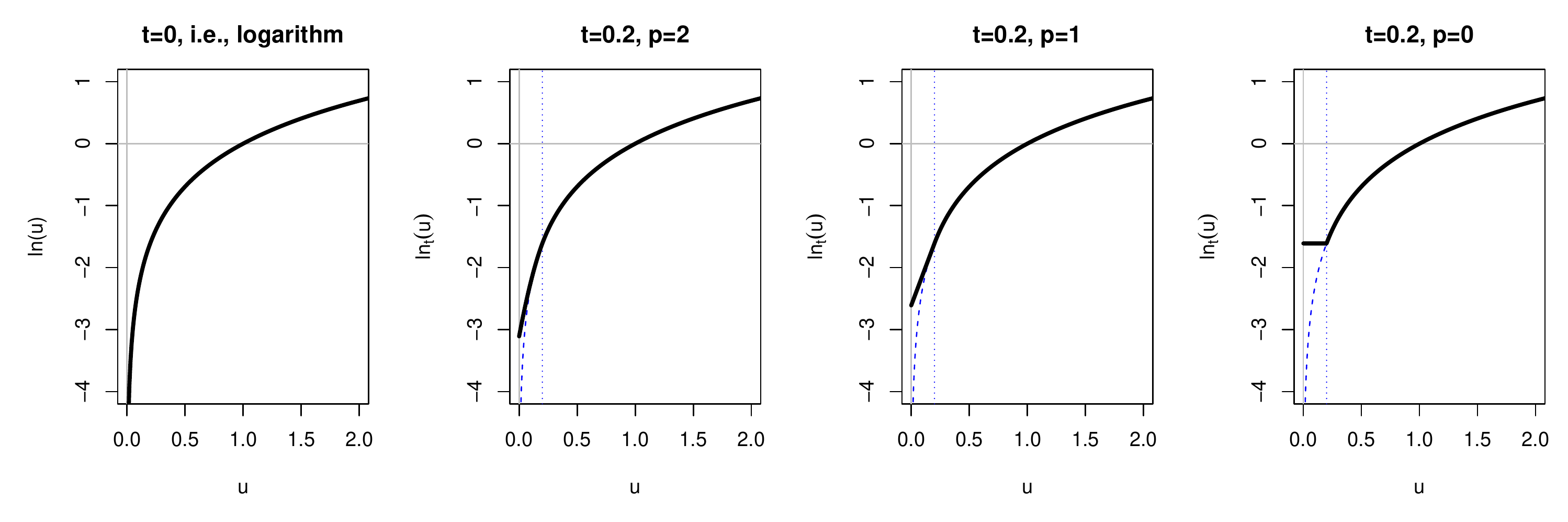}
		\caption{Illustration of $\ln_t(u)$ in bold black with different $p$ and $t$} \label{fig.lnt}
	\end{figure}

	On the other hand, an ideal robust statistical procedure should perform nearly optimally when model assumptions are valid and still maintain good performance when the assumptions are violated.  Motivated by this goal, we  propose the maximum tangent likelihood estimation (MTE) as 
	\begin{align}
	\tilde{\boldsymbol{\beta}} = {\arg\max}_{\boldsymbol{\beta} \in \mathbb{R}^d} \sum_{i=1}^{n} \ln_t (f({\bf z}_i; \boldsymbol{\beta})), \label{eq.MTE}
	\end{align}
	and also propose the penalized maximum tangent likelihood estimation (penalized MTE) for variable selection as
	\begin{align}
	\hat{\boldsymbol{\beta}}=\arg\max_{\boldsymbol{\beta} \in \mathbb{R}^d}\bigg\{ \sum_{i=1}^{n} \ln_t(f(\mathbf{z}_i; \boldsymbol{\beta})) - n\sum_{j=1}^{d} p_{\lambda_{nj}}(|\beta_j|)\bigg\}, \label{eq.PMTE}
	\end{align}
	where the function $\ln_t(\cdot)$ is defined as  
	\begin{align}
	\ln_t(u)=\begin{cases}
	\ln(u) & \text{if } u > t, \\
	\ln(t)+\sum_{k=1}^{p} \frac{\partial^k \ln (v) }{\partial 
	v^k}\big|_{v=t} \frac{(u-t)^k}{k!} & 
	\text{if } 0 \leq u \leq t.
	\end{cases} \label{eq.lnt}
	\end{align}	
	Here $t \geq 0$ is a tuning parameter.
	$\ln_t(u)$ is essentially a $p$-th order Taylor expansion of $\ln(u)$ for 
	$0 \leq u<t$.  Figure \ref{fig.lnt} illustrates the shape of $\ln_t(\cdot)$ 
	with 
	various $p$ and $t$.  Since $\ln_t(u) \to \ln(u)$ as $t \to 0^{+}$, MTE 
	contains the maximum likelihood estimation (MLE) 
	as a special case with $t=0$.  Although $p$ also determines the shape of 
	$\ln_t(\cdot)$, we found out through simulation that its effect is much 
	less significant than that of $t$.  For ease of illustration, throughout 
	this article, we mostly focus $p=1$ (hence the name ``tangent'') unless 
	indicated otherwise.  However, our results are expected to hold for a 
	general $p$.
	
	One advantage of MTE is that, when solving the optimization \eqref{eq.MTE} 
	to obtain $\tilde{\boldsymbol{\beta}}$ (and assuming the regularities 
	conditions in the appendix), we essentially solve a weighted likelihood 
	equation,
	\begin{align}
	0=\sum_{i=1}^{n}\frac{\partial }{\partial \boldsymbol{\beta}} \ln_t (f({\bf z}_i; \boldsymbol{\beta}))=\sum_{i=1}^{n}w_i \frac{\partial }{\partial \boldsymbol{\beta}}\ln (f({\bf z}_i; \boldsymbol{\beta}))   \label{eq.dlnt},
	\end{align}
	where 
	$w_i=[1- (1-f({\bf z}_i; \boldsymbol{\beta})/t )^p 
	]^{\mathbbm{1}\{f({\bf z}_i; \boldsymbol{\beta}) < t\}}$
	and $\mathbbm{1}\{\cdot\}$ is an indicator function.  Note that $t \to 0$, 
	$w_i \to 1$.  In the weighted likelihood equation, the observations that 
	disagree with the assumed model receive low weights.

	Another advantage of MTE is that, when estimating the linear regression 
	coefficients and $p=1$, MTE can be considered 
	as a mixture of minimum Kullback-Leibler (KL) distance estimation and 
	minimum $\ell_2$ distance (L2D) estimation \citep{Lozano2016} or 
	equivalently, 
	exponential squared loss (ESL) estimation \citep{Wang2013}.  
	To see this, we rewrite \eqref{eq.MTE} as
	\begin{align*}
	\tilde{\boldsymbol{\beta}}=\arg\max_{\boldsymbol{\beta} \in 
	\mathbb{R}^d}\bigg\{ \underbrace{\sum_{ i \in \mathcal{A} }\ln( f({\bf 
	z}_i;\boldsymbol{\beta}))}_{\text{minimizing KL}}+ \underbrace{\frac{1}{t} 
	\sum_{i \in \mathcal{A}^c} f({\bf 
	z}_i;\boldsymbol{\beta})}_{\text{minimizing $\ell_2$}} \bigg\},
	\end{align*}
	where $\mathcal{A}=\{i: f({\bf z}_i;\boldsymbol{\beta}) \geq t\}$.  
	Maximizing the first (or second) term alone leads to minimizing the KL 
	distance (or L2D and ESL), respectively.
	Therefore, MTE combines the merits of both, obtains remarkable 
	robustness and still performs well for clean data.
	
	To robustly select variables, we further equip MTE with a penalty 
	$\sum_{j=1}^{d} p_{\lambda_{nj}}(|\beta_j|)$, i.e., the penalized MTE.  We 
	can show that the proposed method is consistent and enjoys oracle 
	property. We also propose a method for adaptively selecting the tuning 
	parameter $t$. In addition, we establish the bound of 
	$\ell_2$ norm of the estimation error under high-dimensional settings.

	Robust variable selection has received increased attention in the recent 
	literature.  In the fixed dimensional setting, \citet{Wang2013} introduced 
	the ESL estimation for robust variable selection.  
	\citet{Wang2007} incorporated the Lasso penalty to least 
	absolute deviation (LAD) estimation for robust linear regression.  
	\citet{Zou2008} proposed composite quantile regression (CQR) for the case 
	where the error variance is infinite. \citet{Alfons2013} 
	considered penalized least trimmed square estimation (LTS).  In the 
	high-dimensional setting, \citet{WangLie2013} considered the properties of
	LAD-Lasso.  
	%\citet{Bradic2011} proposed penalized composite quasi-likelihood 
	% estimation which could be robust to heavy-tailed errors.  
	% \citet{fan2014} proposed penalized quantile regression with weighted  
	% $\ell_1$-penalty for heavy-tailed data.  
	\citet{fan2016huber} studied the penalized Huber's loss for 
	asymmetric 
	contamination.
	\citet{Lozano2016} considered penalized $\ell_2$ distance estimation 
	(L2D) to 
	handle the contamination in the response variable.  
	
	In this paper, we contribute to the literature by 
	proposing a new class of estimators, 
	penalized tangent likelihood estimation,
	and demonstrate its desirable properties in high-dimensional regression 
	estimation and variable selection.
	The proposed class of estimators offers protection for high-dimensional 
	estimation against violation of a particular assumed error distribution.
	It is a generalization of a few existing methods, including penalized least 
	square, L2D, ESL and LTS. 
	Unlike LAD and CQR that essentially estimate the quantile, our approach 
	directly deals with the mean regression \citep{fan2016huber}.  
	Similar to the celebrated Huber loss which is a mixture of least square and 
	LAD, the tangent likelihood could also be considered as a 
	mixture of KL and $\ell_2$ distances.
	However, unlike the Huber loss which is monotone, the proposed 
	class of estimators is essentially penalized redescending M-estimates. 
	We further establish the asymptotic properties in high-dimensional settings 
	for the entire proposed 
	class of estimators, 
	which implies that these aforementioned special cases and our proposed 
	method all enjoy such properties.
	Finally, we demonstrate the advantages of the proposed method through simulations and real data applications.
	
%	Under high-dimensional settings, traditional LASSO penalty is often used in 
%a general loss framework \citep{fan2016huber, Lozano2016}. In this paper, we 
%are able to establish consistency of the penalized MTE with the traditional 
%LASSO penalty when the number of variables $d$ can grow exponentially with the 
%sample size $n$. In Section \ref{sec.MTE_linear_regression}, we show that the 
%MTE can be viewed as a mixture of minimum KL distance estimation and minimum 
%$\ell_2$ distance estimation, where $t$ is a robustness tuning parameter 
%balancing the trade-off of efficiency and robustness. Equivalently, the 
%penalized MTE can be regarded as a class of estimators that include the 
%penalized MLE and minimum $\ell_2$ distance robust estimator as special cases. 
%The consistency results we establish imply that for any $t>0$, the proposed 
%penalized MTE can achieve consistency in the order of $\sqrt{\ln(d)/n}$, the 
%optimal convergence rate.  
	 	
	The paper is organized as follows.  In Section \ref{sec.MTEprop}, we 
	formally 
	introduce MTE, study its properties 
	and discuss its links to other estimators.  In Section \ref{sec.asymPTE}, 
	we further introduce the penalized MTE for variable selection, and 
	demonstrate its asymptotic properties through an 
	analysis of consistency, oracle property.
	We discuss the implementation aspect of the 
	method such as selection of tuning parameters in Section 
	\ref{sec.algo} and present numerical results in Section 
	\ref{sec.numer}.  Finally, we conclude with a discussion in Section 
	\ref{sec.concl} and relegate the proofs to the supplementary materials.
	
	%\citep{Tibshirani1996,Fan2001,Zou2006,Wang2007,Zou2008,Wu2009,Song2010,Wang2013,fan2004,ZhaoYu2006,meinshausen2006,FanLi2006,candes2007,ZhangHuang2008,Huang2008,zhang2010,FanLv2010,Bradic2011,Belloni2011,fan2014,ZhangHuang2008,Huang2008,zhang2010,Bradic2011,Belloni2011}

	%\newpage
	%%%%%%%%%%%%%%%%%%%%%%%%%%%%%%%%%%%%%%%%%%%%%%%%%%%%%%%%%%%%%%%%%%%%%%%%%%%%%%%%%%%%%%%%%%%%%%%%
	\section{Maximum Tangent Likelihood Estimation}\label{sec.MTEprop}
	
	%%%%%%%%%%%%%%%%%%%%%%%%%%%%%%%%%%%%%%%%%%%%%%%%%%%%%%%%%%%%%%%%%%%%%%%%%%%%%%%%%%%%%%%%%%%%%%%%
	\subsection{Motivations of Maximum Tangent Likelihood Estimation}\label{sec:MTE_general}
	
	Let $({\bf z}_1, \dotso,{\bf z}_n)$ be an i.i.d. random sample from a general probability 
	model $f({\bf z}; \boldsymbol{\beta})$ with parameter $\boldsymbol{\beta} 
	\in \mathcal{B} \subset \mathbb{R}^d$.  We define the maximum tangent 
	likelihood estimator (MTE) of $\boldsymbol{\beta}$ as in 	
	\eqref{eq.MTE}.  
	Assuming 
	regularity conditions, we can solve $\tilde{\boldsymbol{\beta}}$ as the 
	root of the tangent likelihood equation \eqref{eq.dlnt}, in which 
	observations that disagree with the assumed model are downweighted.  To 
	solve the weighted likelihood equation, we iterate the procedures of 
	solving the parameter given the weights and updating the weights with new 
	parameter (iterative re-weighted algorithm).
	
	When $p=1$, the weight simplifies to $w_i=\min\{1, f({\bf z}_i, \boldsymbol{\beta})/t \}$, hence the tangent likelihood equation becomes
	\begin{align*}
	0&=\sum_{i=1}^{n}\left[\frac{\partial}{\partial \boldsymbol{\beta}} \ln (f({\bf z}_i; \boldsymbol{\beta})) \right] \min\left\{1, \frac{f({\bf z}_i, \boldsymbol{\beta})}{t} \right\},
	\end{align*}
	which is a Mallows type estimator \citep{Mallows1975}.  So if the observation has a likelihood below $t$, it is assigned partial weight, $f({\bf z}_i, \boldsymbol{\beta})/t$.  Otherwise, the observation is assigned full weight.  When estimating the mean of a normal distribution, we have $\tilde{\mu}=(\sum_{i=1}^{n} w_i {\bf z}_i)/\sum_{i=1}^{n}w_i$ where $w_i= \min(1,\varphi({\bf z}_i;\tilde{\mu},\tilde{\sigma^2})/t)$ and $\varphi(\cdot)$ is the Gaussian density function.  $\tilde{\mu}$ is essentially a weighted mean.
	
	When $p=0$, we have $w_i=\mathbbm{1}\{f({\bf z}_i; \boldsymbol{\beta}) \geq t\}$
	and the tangent likelihood equation becomes
	\begin{align*}
	0=\sum_{i=1}^{n}\left[\frac{\partial}{\partial \boldsymbol{\beta}} \ln (f({\bf z}_i; \boldsymbol{\beta}))\right] \mathbbm{1}\{f({\bf z}_i; \boldsymbol{\beta}) \geq t\}=\sum_{i \in \mathcal{A}}\frac{\partial }{\partial \boldsymbol{\beta}}\ln (f({\bf z}_i; \boldsymbol{\beta})),
	\end{align*}
	where $\mathcal{A}=\{i: f({\bf z}_i; \boldsymbol{\beta}) \geq t \}$.  That is, we completely discard the data points whose likelihoods are below $t$.  This follows similar spirit as in the trimmed likelihood/least square estimation proposed by \citet{Hadi1997} and \citet{Alfons2013}.  When estimating the mean of a normal distribution, we have
	$\tilde{\mu}=(\sum_{i \in \mathcal{A}} {\bf z}_i)/|\mathcal{A}|$ where 
	$\mathcal{A}=\{i: \varphi({\bf z}_i;\tilde{\mu},\tilde{\sigma^2}) \geq t 
	\}$, i.e., a trimmed mean with data points whose likelihoods below $t$ are 
	removed.  MTE may be also related with an early work by \citet{Field1994} 
	and the empirical likelihood estimation \citep{Owen2001book}.  Next, we 
	focus our attention to applying MTE to linear models and variable selection 
	through penalization under both fixed and high-dimensional settings.

	%%%%%%%%%%%%%%%%%%%%%%%%%%%%%%%%%%%%%%%%%%%%%%%%%%%%%%%%%%%%%%%%%%%%%%%%%%%%%%%%%%%%%%%%%%%%%%%%
	\subsection{Motivations of MTE for Linear Regressions and its Connections 
	to Other Estimators}\label{sec.MTE_linear_regression}
	%%%%%%%%%%%%%%%%%%%%%%%%%%%%%%%%%%%%%%%%%%%%%%%%%%%%%%%%%%%%%%%%%%%%%%%%%%%%%%%%%%%%%%%%%%%%%%%%
	We apply MTE to linear models.  Consider a linear regression model
	\begin{align}
	y_i=\mathbf{x}_i^T \boldsymbol{\beta} + \epsilon_i, \qquad i=1,\dotso, n, 
	\label{eq.lreg}
	\end{align}
	where ${\bf z}_i=(y_i, \mathbf{x}^T_i)$ is the $i$th observation.  $y_i \in 
	\mathbb{R}, \mathbf{x}_i \in \mathbb{R}^d$. $\boldsymbol{\beta}=(\beta_{1}, 
	\dotso, \beta_{d}) \in \mathbb{R}^d$ is an unknown regression coefficient 
	vector.  $\epsilon_i$ is an i.i.d. random error that is independent from 
	$\mathbf{x}_i$. We assume that the random error $\epsilon_i$ follows a 
	symmetric parametric distribution $f(\cdot)$ with zero mean and constant 
	variance $\sigma^2$, which can be considered as a nuisance parameter 
	\citep{godambe1974} and is usually estimated by a high breakdown point 
	preliminary scale estimate $\sigma^{2}_R$, such as LAD, L2D, and LTS 
	\citep{Huber2009,Maronna2006,Hampel1986,vanderVaart1998}.    
	Throughout this article, we assume $f(\cdot)$ to be a Gaussian probability 
	density function with zero mean. However, it is expected that the 
	methodology presented in this article to hold for a wide range of densities 
	well beyond the Gaussian density.  
	
	Let us show how MTE for linear regression is related to the minimum KL 
	distance estimation and the minimum $\ell_2$ distance estimation.  We start 
	by rewriting \eqref{eq.MTE} for $p=1$ as 
	\begin{align}
	\tilde{\boldsymbol{\beta}}=\arg \max_{\boldsymbol{\beta} \in \mathbb{R}^d} 
	\left\{ \sum_{ i \in \mathcal{A}}\ln( f({\bf z}_i;\boldsymbol{\beta}))+ 
	\frac{1}{t} \sum_{i \in \mathcal{A}^c} f({\bf z}_i;\boldsymbol{\beta}) 
	\right\}, \label{eq.MTEregEqv}
	\end{align}
	where $\mathcal{A}=\{i:f({\bf z}_i;\boldsymbol{\beta})\geq t \}$.
	
	First, note that the minimum KL distance estimate 
	$\tilde{\boldsymbol{\beta}}_{\textup{KL}}$ is essentially the MLE, that is
	\begin{align}\label{eq.MiniKL}
	\tilde{\boldsymbol{\beta}}_{\textup{KL}} &=\arg \max_{\boldsymbol{\beta} 
	\in \mathbb{R}^d} \left\{ \sum_{ i \in \mathcal{A}} \ln(f({\bf 
	z}_i;\boldsymbol{\beta}))+ \sum_{i \in \mathcal{A}^c} \ln(f({\bf 
	z}_i;\boldsymbol{\beta})) \right\}.
	\end{align}

	Second, note that the minimum $\ell_2$ distance estimate 
	$\tilde{\boldsymbol{\beta}}_{\ell_2}$ for linear regression is 
	\citep{Scott2001,Lozano2016}
	\begin{align}\label{eq.MiniL2}
	\tilde{\boldsymbol{\beta}}_{\ell_2}=\arg \max_{\boldsymbol{\beta} \in 
	\mathbb{R}^d} \left\{ \sum_{ i \in \mathcal{A}} f({\bf 
	z}_i;\boldsymbol{\beta})+ \sum_{i \in \mathcal{A}^c} f({\bf 
	z}_i;\boldsymbol{\beta}) \right\}.
	\end{align}

	\begin{remark}
	To understand \eqref{eq.MiniL2}, consider the $\ell_2$ distance between the 
	parametric distribution of $y$ given ${\bf x}$, $p(y|{\bf 
	x},\boldsymbol{\beta})$, and the true distribution of $y$ given ${\bf x}$, 
	$p(y|{\bf x})$,
	\begin{align*}
	\int (p(y|{\bf x},\boldsymbol{\beta})-p(y|{\bf x}))^2 dy &= \int p(y|{\bf 
	x},\boldsymbol{\beta})^2 dy+\int p(y|{\bf x})^2dy\\
	& \quad -2\int p(y|{\bf x},\boldsymbol{\beta})p(y|{\bf x}) dy\\
	&= \int p(y|{\bf x},\boldsymbol{\beta})^2 dy+\int p(y|{\bf 
	x})^2dy-2\mathbb{E} f(y-{\bf x}^T\boldsymbol{\beta}).
	\end{align*}
	For linear regressions, $\int p(y|{\bf x},\boldsymbol{\beta})^2 dy=\int f(y 
	- {\bf x}^T\boldsymbol{\beta})^2 dy$ does not depend on 
	$\boldsymbol{\beta}$. Hence, minimizing the $\ell_2$ distance with respect 
	to $\boldsymbol{\beta}$ is equivalent to maximizing $\mathbb{E} f(y-{\bf 
	x}^T\boldsymbol{\beta})$.  When observing a sample, we replace $\mathbb{E} 
	f(y-{\bf x}^T\boldsymbol{\beta})$ with its empirical mean $\sum_{i=1}^n 
	f({\bf z}_i;\boldsymbol{\beta})/n$, and obtain 
	$\tilde{\boldsymbol{\beta}}_{\ell_2}=\arg \max_{\boldsymbol{\beta} \in 
	\mathbb{R}^d} \sum_{i=1}^{n} f({\bf z}_i;\boldsymbol{\beta})$.  
	\end{remark}	

	Comparing \eqref{eq.MTEregEqv} with \eqref{eq.MiniKL} and 
	\eqref{eq.MiniL2}, we understand that MTE can be considered as minimizing a 
	mixture of KL and $\ell_2$ distances.  When $t=0$, all the observations 
	fall into the set $\mathcal{A}$, and MTE becomes the minimum KL distance 
	estimation.  As $t$ gradually increases away from 0, some observations with 
	relatively low likelihoods gradually move from $\mathcal{A}$ to 
	$\mathcal{A}^c$.  When $t$ is sufficiently large, all observations have 
	moved from $\mathcal{A}$ to $\mathcal{A}^c$, and MTE becomes the minimum 
	$\ell_2$ distance estimation.  
	
	With an appropriately selected $t$, we have observations in both 
	$\mathcal{A}$ and $\mathcal{A}^c$.  The observations in $\mathcal{A}^c$ are 
	the potential outliers.  If they were to be used in the pure minimum KL 
	distance estimation, we would have an unstable estimate.  Meanwhile, the 
	observations in $\mathcal{A}$ are the good observations.  If they were to 
	be used in the pure minimum $\ell_2$ distance estimation, we would have an 
	inefficient estimate.  Therefore, MTE minimizes the KL distance for the 
	observations in $\mathcal{A}$ and minimizes the $\ell_2$ distance for the 
	observations in $\mathcal{A}^c$ to preserve efficiency and gain 
	robustness.  
	
	Finally, we summarize the links between MTE and other estimators for linear 
	regression as special cases.  Suppose $T$ is a sufficiently large number. 
	When $0<t<T$ and $p=0$, MTE is asymptotically equivalent to LTS 
	\citep{Hadi1997,Alfons2013}.  When $0<t<T$ and $p=1$, MTE can be considered 
	as a mixture of minimum KL distance and minimum $\ell_2$ distance.  When $t 
	\geq T$ and $p=1$, MTE is equivalent to L2D or ESL.  Lastly, when $t=0$ or 
	when $p=+\infty$, MTE is essentially MLE 
	or minimum KL distance estimation.
	
%	\begin{table}[h!]
%		\begin{center}
%			\begin{tabular}{| c | c | p {5cm} | p {5cm} |}
%				\hline
%				& $t=0$ & $0<t<T$ & $t\geq T$ \\ \hline
%				$p=0$ & MLE & trimmed least square estimation 
%\citep{Hadi1997,Alfons2013} & not applicable\\
%				\hline
%				$p=1$ &	MLE & mixture of minimum KL distance and minimum 
%$\ell_2$ distance & minimum $\ell_2$ distance \citep{Scott2001,Lozano2016}\\ 
%				\hline
%				$p=2,3,...$ & MLE & redescending M-estimate & redescending 
%M-estimate \\
%				\hline
%				$p=+\infty$ & MLE & MLE & MLE \\
%				\hline
%			\end{tabular}
%			\caption{Summary of the relationship between MTE and other 
%estimators for linear regressions. $T>0$ is a sufficiently large number.}
%			\label{table.summary}
%		\end{center}
%	\end{table}
	
	%An alternative way to understand MTE is that, the minimum $\ell_2$ 
	%distance estimation is equivalently to maximize $\ln \big(\sum_{i=1}^{n} 
	%f({\bf z}_i;\boldsymbol{\beta})\big)$.  The minimum KL distance estimation 
	%is to maximize $\sum_{i=1}^{n} \ln (f({\bf z}_i;\boldsymbol{\beta}))$.  
	%The 
	%distinction between these two is whether the $\ln(\cdot)$ is inside or 
	%outside of the summation.  On the other hand, MTE only brings parts of the 
	%$\ln(\cdot)$ functions outside the summation sign and keep most of 
	%$\ln(\cdot)$ inside the summation.  

	%%%%%%%%%%%%%%%%%%%%%%%%%%%%%%%%%%%%%%%%%%%%%%%%%%%%%%%%%%%%%%%%%%%%%%%%%%%%%%%%%%%%%%%%%%%%%%%%
	\subsection{Asymptotic Properties of Maximum Tangent Likelihood Estimation}\label{sec.MTE_properties}
	%%%%%%%%%%%%%%%%%%%%%%%%%%%%%%%%%%%%%%%%%%%%%%%%%%%%%%%%%%%%%%%%%%%%%%%%%%%%%%%%%%%%%%%%%%%%%%%%
	We present asymptotic properties of MTE.
	First define $\boldsymbol{\beta}^*_t = \arg\max_{\boldsymbol{\beta} \in 
	\mathcal{B}} \mathbb{E}_{\boldsymbol{\beta}_0} \ln_t (f({\bf z}; 
	\boldsymbol{\beta}))$ where $\boldsymbol{\beta}_0$ is the true parameter 
	and $t \geq 0$.  
	
	\begin{theorem}{\label{thm.MTEconsistency}}
		Under the regularity conditions specified in the supplementary materials, with probability going to 1, there exists a unique solution $\tilde{\boldsymbol{\beta}}$ for equation \eqref{eq.MTE}. Furthermore, we have $\tilde{\boldsymbol{\beta}} \overset{p} \to \boldsymbol{\beta}^*_t$ as $n \to \infty$.
	\end{theorem}
	
	\begin{theorem}{\label{thm.MTEefficiency}}
		Under the regularity conditions specified in the supplementary materials, we have 
		\begin{align*}
		\sqrt{n}{\bf \Omega}^{-1/2}(\tilde{\boldsymbol{\beta}} - \boldsymbol{\beta}^*_t) \overset{d}\to N({\bf 0}, {\bf I})  \quad \text{as} \quad n \to \infty, 
		\end{align*}	
		where ${\bf I}$ is a $d \times d$ identity matrix, ${\bf \Omega} = {\bf A}^{-1} {\bf B} {\bf A}^{-1}$, ${\bf A} = \partial^2  \mathbb{E}_{\boldsymbol{\beta}_0} \big[ \ln_t(f({\bf z};\boldsymbol{\beta}_t^*)) \big] /\partial \boldsymbol{\beta}^2$, and ${\bf B} = \mathbb{E}_{\boldsymbol{\beta}_0} \left[ (\partial \ln_t(f({\bf z};\boldsymbol{\beta}_t^*))/\partial \boldsymbol{\beta} )(\partial \ln_t(f({\bf z};\boldsymbol{\beta}_t^*))/\partial \boldsymbol{\beta})^T  \right]$.  When $t \to 0^+$, we have $\boldsymbol{\beta}^*_t \to \boldsymbol{\beta}_0$ and ${\bf \Omega}$ becomes the inverse of Fisher information matrix.
	\end{theorem}
	In general, $\boldsymbol{\beta}^*_t$ is not necessarily the same as 
	$\boldsymbol{\beta}_0$ for $t>0$.  However, when $\boldsymbol{\beta}_0$ 
	represents the location parameter of a symmetric distribution such as 
	linear regression coefficients, then we have 
	$\boldsymbol{\beta}^*_t=\boldsymbol{\beta}_0$, which means MTE is indeed a 
	consistent estimator and has asymptotic normality for such a case.

	\begin{theorem}[Consistency and asymptotic normality]\label{thm.MTE_Regression_consistency}
		Under the regularity conditions specified in the supplementary materials, for linear regression $y_i=\mathbf{x}_i^T \boldsymbol{\beta}_0 + \epsilon_i$, suppose the error $\epsilon_i$ follows a symmetric distribution with zero mean.  Then we have $\boldsymbol{\beta}^*_t=\boldsymbol{\beta}_0$ for any $t>0$.  That is, MTE of the regression coefficient $\tilde{\boldsymbol{\beta}}$ defined in equation \eqref{eq.MTE} is consistent and asymptotically normal for any $t > 0$.
	\end{theorem}
	
	With a consistent MTE, we can further apply it into variable selection problem for linear regression and study its properties.
	
	%%%%%%%%%%%%%%%%%%%%%%%%%%%%%%%%%%%%%%%%%%%%%%%%%%%%%%%%%%%%%%%%%%%%%%%%%%%%%%%%%%%%%%%%%%%%%%%%
	\section{Penalized MTE for Variable Selection}\label{sec.asymPTE}
		
	Usually, some of the elements of $\boldsymbol{\beta}_0=(\beta_{01}, \dotso, 
	\beta_{0d})$ in the linear regression are zeros, meaning that the 
	corresponding covariates are not affecting $y_i$.  It is a fundamental task 
	to build a linear regression model with important covariates and estimate 
	their coefficients.  Without loss of generality, assume 
	$\boldsymbol{\beta}=(\boldsymbol{\beta}^T_{S}, 
	\boldsymbol{\beta}^T_{S^c})^T$ where $S=\left\{j: \beta_{0j} \neq 0, 
	j=1,\dotso, d\right\}=\left\{1,\dotso, s\right\}$ and $|S|=s$, 
	$\boldsymbol{\beta}_S \in \mathbb{R}^s$ and $\boldsymbol{\beta}_{S^c} \in 
	\mathbb{R}^{d-s}$.  The true regression coefficient is 
	$\boldsymbol{\beta}_0=(\boldsymbol{\beta}^T_{0S}, 
	\boldsymbol{\beta}^T_{0S^c})^T$ where all elements in 
	$\boldsymbol{\beta}_{0S}$ are non-zeros and all elements in 
	$\boldsymbol{\beta}_{0S^c}$ are zeros.  
	To perform variable selection and coefficient estimation simultaneously, we 
	use the penalized MTE, $\hat{\boldsymbol{\beta}}$, defined in 
	\eqref{eq.PMTE}.
	
	\subsection{\texorpdfstring{Asymptotic Properties with Fixed 
	Dimensionality}{Asymptotic Properties with Fixed 
	Dimensionality}}\label{sec.asym_fixed_d}
	
	When the number of covariates $d$ is fixed and the sample size $n \to 
	\infty$, the penalized MTE is $\sqrt{n}$-consistent and enjoys the oracle 
	property under mild regularity conditions.  Let $a_n  = \max \{ 
	p'_{\lambda_{nj}} (\left| \beta_{0j} \right|) : \beta_{0j} \neq 0 \}$ and 
	$b_n = \max \{ p''_{\lambda_{nj}} (\left| \beta_{0j} \right|) : \beta_{0j} 
	\neq 0 \}$.  We provide following theorems.
	\begin{theorem}[$\sqrt{n}$-consistency]{\label{thm.PMTECons}}
		Under the regularity conditions specified in the supplementary materials, suppose $a_n=O_p(n^{-1/2})$, $b_n=o_p(1)$ and $t > 0$, then there exists a local maximizer $\hat{\boldsymbol{\beta}}$, such that $\|\hat{\boldsymbol{\beta}}-\boldsymbol{\beta}_0\|_2 = O_p(n^{-1/2})$.
	\end{theorem}
	
	\begin{theorem}[Oracle property]{\label{thm.oracle}} 
		Assume that the penalty function satisfies 
		\begin{align}
			\liminf_{n \to \infty} \liminf_{\theta \to 0+} \left\{ \min_{s+1\leq j \leq d} p'_{\lambda_{nj}} (\theta) / \lambda_{nj} \right\} >0, \label{eq.penaltyfororacle}
		\end{align}
		and the regularization parameter $\lambda_{nj}$ satisfies
		\begin{align}
			\max_{1\leq j \leq s}(\sqrt{n}\lambda_{nj})=o_p(1) \quad \text{and} \quad 1/\min_{s+1 \leq j \leq d}(\sqrt{n}\lambda_{nj})=o_p(1). \label{eq.lambdacondt}
		\end{align}
		Suppose $t > 0$, then $\hat{\boldsymbol{\beta}}$ satisfies:
		
		\quad (a) Sparsity: $\hat{\boldsymbol{\beta}}_{S^c} = \mathbf{0}$ with probability 1;
		
		\quad (b) Asymptotic normality for $\hat{\boldsymbol{\beta}}_{S}$:	
		\[
		\sqrt{n}({\bf J}_S + {\bf \Sigma}_1) \left\{ \hat{\boldsymbol{\beta}}_{S} - \boldsymbol{\beta}_{0S} + ({\bf J}_S + {\bf \Sigma}_1)^{-1} {\bf b} \right\} \xrightarrow{d} N (\mathbf{0}, {\bf \Sigma}_2 ), 
		\]
		where 
	${\bf \Sigma}_1 = \textup{diag} ( p''_{\lambda_{n1}} (\left| \beta_{01} 
	\right|), \dotso, p''_{\lambda_{ns}} (\left| \beta_{0s} \right|))$, ${\bf 
	\Sigma}_2 = \textup{cov} [ \partial \ln_{t}( f({\bf 
	z};\boldsymbol{\beta}_0)) /\partial \boldsymbol{\beta}_S]$, 
	
	\noindent ${\bf J}_S = \mathbb{E}[ \partial^2\ln_{t}( f({\bf 
	z};\boldsymbol{\beta}_0)) / \partial{\boldsymbol{\beta}_S}\partial 
	\boldsymbol{\beta}^T_S]$, and  ${\bf b} = ( p'_{\lambda_{n1}}(\left| 
	\beta_{01} \right|) \textup{sgn} (\beta_{01}), \dotso, p'_{\lambda_{ns}}(\left| \beta_{0s} \right|) \textup{sgn} (\beta_{0s}) )^T$.
	\end{theorem}
	
	By Theorem \ref{thm.oracle}, it is straightforward to derive the asymptotic covariance matrix for $\hat{\boldsymbol{\beta}}_S$,
	\begin{align}
		\text{Var}(\hat{\boldsymbol{\beta}}_S)=\frac{1}{n} \{{\bf J}_S + {\bf \Sigma}_1 \}^{-1} {\bf \Sigma}_2 \{{\bf J}_S + {\bf \Sigma}_1 \}^{-1}. \label{eq.asycovmat}
	\end{align}	
	We use this analytical form of the variance-covariance matrix of 
	$\hat{\boldsymbol{\beta}}_S$ in the choice of tuning parameter $t$ (as 
	detailed in Section \ref{subsec.tuningt}).
	It is easy to see that penalty functions such as adaptive-Lasso \citep{Zou2006} satisfy conditions \eqref{eq.penaltyfororacle} and \eqref{eq.lambdacondt} to achieve the oracle property, unlike the traditional Lasso penalty. Nonetheless, the penalized MTE with the traditional Lasso penalty can still achieve consistency.

	\subsection{\texorpdfstring{Consistency under High-Dimensional Regression}{Consistency under High-Dimensional Regression}}\label{sec.asym_ultrahigh}
	
	We further consider the penalized MTE for modern high-dimensional linear regression setting, where the number of covariates $d$ is allowed to approach infinity as well as the sample size $n$ in model \eqref{eq.lreg}. In particular, we consider $\ln (d)/n \to 0$ as $n \to \infty$ and $d \to \infty$. In this setting, the true coefficient vector $\boldsymbol{\beta}_0$ is usually assumed to be sparse. Regularization method with $\ell_1$ penalty is among the popular methods to achieve sparse estimation. In this section, we establish the statistical consistency of MTE with Lasso penalty function, termed as \textit{MTE-Lasso}, by establishing the $\ell_2$-norm bound $\|\hat{\boldsymbol{\beta}}-\boldsymbol{\beta}_0\|_2$. Specifically, we consider the estimator
	\begin{align}
	\hat{\boldsymbol{\beta}}=\arg\min_{\boldsymbol{\beta} \in \mathbb{R}^d}\bigg\{ \mathcal{L}(\boldsymbol{\beta}) + \lambda_{n}  \sum_{j=1}^{d} |\beta_j|\bigg\}, \label{eq.HDPMTE}
	\end{align}
	where $ \mathcal{L}(\boldsymbol{\beta}) = -(1/n) \sum_{i=1}^{n} \ln_{t}(f({\bf z}_i;\boldsymbol{\beta}))$ is MTE loss function, and $\lambda_n$ is the regularization parameter of $\ell_1$ penalty. Let $\hat{\bf \Delta}=\hat{\boldsymbol{\beta}}-\boldsymbol{\beta}_0$ and define $\mathbb{C}(S)=\{{\bf \Delta} \in \mathbb{R}^d:3\|{\bf \Delta}_{S}\|_1 \geq \| {\bf \Delta}_{S^c}\|_1\}$ where ${\bf \Delta}_{S}$ and ${\bf \Delta}_{S^c}$ are the projections of ${\bf \Delta}$ onto the coordinate sets $S$ and $S^c$ respectively.  We further have the following assumptions. 	
	
	\begin{enumerate}[label={A\arabic*}, leftmargin=*, start=1]
		\item The regressors are bounded, i.e., $\|{\bf x}_i\|_{\infty}=M<+\infty$ for all $i=1,\dotso,n$.
		\item The design matrix ${\bf X} = ({\bf x}_1, ...,{\bf x}_n)^T$ satisfies the restricted eigenvalue condition, $\|{\bf X}{\bf \Delta}\|_2^2/n \geq \kappa_{\textrm{RE}}\|{\bf \Delta}\|_2^2\textrm{, for all }{\bf \Delta} \in \mathbb{C}(S)$ where $\kappa_{\textrm{RE}}>0$.
		%\item For any $\boldsymbol{\nu} \in \mathbb{R}^d$, and ${\bf x}_i, i=1, \dotso, n$, the random variable ${\bf x}_i^T \boldsymbol{\nu}$ is sub-Gaussian with parameter at most $\eta^2 \|\boldsymbol{\nu}\|^2_2$.
	\end{enumerate}
	
	Note that the assumptions above are also imposed in \citet{Lozano2016}. 
	%However, we do not require the error distribution to be symmetric as in most high-dimensional regression theories. 
	In order to establish the bound for $\|\hat{\boldsymbol{\beta}}-\boldsymbol{\beta}_0\|_2$ in high-dimensional regressions, we need to verify two critical conditions: (1) the boundedness of the gradient of the loss function $\mathcal{L}$ at the true parameter $\boldsymbol{\beta}_0$ and (2) the restricted strong convexity (RSC) condition of the loss function $\mathcal{L}$ in the neighborhood of the true parameter $\boldsymbol{\beta}_0$.  
	
	We show that the first condition holds with high probability in the following Lemma.
	\begin{lemma}{\label{lem.highdimlemma1}}
		Under Assumption 1, for $t>0$, we have
		\begin{align*}
		P\bigg(\left\|\frac{\partial\mathcal{L}(\boldsymbol{\beta}_0)}{\partial \boldsymbol{\beta}} \right\|_{\infty} \leq \xi \sqrt{\frac{\ln (d)}{n}}\bigg) \geq 1-2\exp(-\alpha_1 \ln (d) ),
		\end{align*}
		where $\alpha_1>0$ is a constant, $\xi=C_t\sqrt{2(\alpha_1+1)}$ and 
		$C_t=Mf(\sigma_R)/(t\sigma_R)$.
		
	\end{lemma}
	
	Lemma \ref{lem.highdimlemma1} shows that $\partial 
	\mathcal{L}(\boldsymbol{\beta}_0)/\partial \boldsymbol{\beta}$ is bounded 
	with high probability and also provides the form of the bound.  This bound 
	plays an important role in deciding the convergence rate of 
	$\hat{\boldsymbol{\beta}}$ as shown in Theorem \ref{thm.highdimth1}.  Since 
	$f$ represents the normal density function, when $t$ increases, $C_t$ 
	decreases, hence the bound also decreases.  It implies that the surface of 
	the loss function around the true parameter $\boldsymbol{\beta}_0$ becomes 
	flatter as $t$ becomes larger.  Lemma \ref{lem.highdimlemma1} corresponds to the
	sub-Gaussian tail condition, which ensures the boundedness of gradient of 
	least square loss \citep{Negahban2012}. 
	The proof is 
	given in the supplementary materials.  In the proof, we particularly discuss the normal 
	density case and give the form of $C_t$.
	
	It is understood that the estimation error $\hat{\bf \Delta}$ belongs to $\mathbb{C}(S)$ when the regularization parameter $\lambda_n \geq 2\|\partial \mathcal{L}(\boldsymbol{\beta}_0)/\partial \boldsymbol{\beta}\|_{\infty}$ \citep[Lemma 1, p.543-544]{Negahban2012}.  Therefore, Lemma \ref{lem.highdimlemma1} suggests that we could choose the regularization parameter $\lambda_n=2\xi\sqrt{\ln(d)/n}$ in the penalized MTE to force $\hat{\bf \Delta} \in \mathbb{C}(S)$.  Such a choice of $\lambda_n$ is valid with probability at least $1-2\exp(-\alpha_2n\lambda_n^2)$ where $\alpha_2=\alpha_1/(4\xi^2)$.  
	
	Given that $\hat{\bf \Delta} \in \mathbb{C}(S)$, we next verify the RSC condition of the loss function $\mathcal{L}$ to establish the estimation error bound.  Before showing the result, we provide the definition of RSC.
	\begin{definition}[Restricted strong convexity]\label{def.RSC}
		The loss function $\mathcal{L}$ satisfies restricted strong convexity (RSC) with curvature $\kappa_1>0$ and tolerance $\tau$ over the set $\mathbb{C}(S)$ if  $\mathcal{L}(\boldsymbol{\beta}_0+{\bf \Delta}) - \mathcal{L}(\boldsymbol{\beta}_0) - [\partial\mathcal{L}(\boldsymbol{\beta}_0)/\partial \boldsymbol{\beta}]^T {\bf \Delta}\geq \kappa_1 \|{\bf \Delta}\|_2^2 + \tau^2$ for all ${\bf \Delta} \in \mathbb{C}(S)$.
	\end{definition}
	
	\begin{lemma}{\label{lem.highdimlemma2}}
		Assume that the random error $\epsilon$ satisfies the tail condition
		\begin{align*}
		P\left(|\epsilon|>\sqrt{c_0 R}-4\sqrt{s}Mu\right)= \kappa_u \leq \Big(1+\frac{c_0}{c_1}2e^{-3/2}\Big)^{-1},
		\end{align*}
		where $c_0=\sigma^2_R$, $c_1=c_0^{3/2}t\sqrt{2\pi}$, and $R=-2\ln(t\sqrt{2\pi c_0})$ with tuning parameter $t$.  Under Assumptions A1 and A2, consider the set $\mathbb{H}(S,u)=\{{\bf \Delta} \in \mathbb{C}(S): \|{\bf \Delta}\|_2=u\}$, for any $u < \sqrt{c_0R}/(4M\sqrt{s})$, and $\mathbf{\Delta} \in \mathbb{H}(S,u)$, it holds that
		\begin{align*}
		\mathcal{L} (\boldsymbol{\beta}_0+{\bf \Delta})-\mathcal{L}(\boldsymbol{\beta}_0) -\bigg(\frac{\partial\mathcal{L}(\boldsymbol{\beta}_0)}{\partial \boldsymbol{\beta}}\bigg)^T {\bf \Delta} \geq  \kappa_1\|{\bf \Delta}\|_2(\|{\bf \Delta}\|_2-\kappa_2\sqrt{\frac{\ln (d)}{n}} \|{\bf \Delta}\|_1)
		\end{align*}
		with probability at least $1-\alpha_3\exp(-\alpha_4 n)$ for some positive constants $\alpha_3$ and $\alpha_4$, where $\kappa_1=\left(1/c_0-c_2\kappa_u\right) \kappa_{\textup{RE}}/4$, $\kappa_2=97 c_2 M^2\sqrt{s}/ (2 \kappa_1)$, and $c_2=1/c_0+2e^{-3/2}/c_1$.
	\end{lemma}
	As we can see, the curvature of the loss function within the neighborhood 
	of $\boldsymbol{\beta}_0$ in the direction of $\mathbb{C}(S)$ is measured 
	by $\kappa_1$.  It can be shown that this curvature increases as $t$ 
	decreases to 0.  In particular, when $t$ decreases to 0, $R$ increases to 
	$+\infty$, and $\kappa_u$ decreases to 0.  Furthermore, for most of the distributions of $\epsilon$, it is 
	straightforward to show that when $t$ decreases to 0, $\kappa_1$ increases 
	to $\kappa_{\textup{RE}}/(4c_0)$.  It implies that as $t$ decreases, the 
	surface of the loss function become more convex which leads to a better 
	convergence rate.
	
	Note that since $\boldsymbol{\Delta} \in \mathbb{C}(S)$, we have $\|\boldsymbol{\Delta}\|_1 \leq 4\|\boldsymbol{\Delta}_S\|_1 \leq 4\sqrt{s} \|\boldsymbol{\Delta}\|_2$, therefore, the results of Lemma \ref{lem.highdimlemma2} becomes
	\begin{align*}
		\mathcal{L} (\boldsymbol{\beta}_0+{\bf \Delta})-\mathcal{L}(\boldsymbol{\beta}_0) -\bigg(\frac{\partial\mathcal{L}(\boldsymbol{\beta}_0)}{\partial \boldsymbol{\beta}}\bigg)^T {\bf \Delta} \geq  \frac{\kappa_1}{2} \|{\bf \Delta}\|^2_2,
	\end{align*}
	when $n > 64 \kappa_2^2 s \ln (d)$.

	With the results provided by Lemmas \ref{lem.highdimlemma1} and \ref{lem.highdimlemma2}, we are ready to establish the bound for $\ell_2$ norm of the estimation error.
	
	\begin{theorem}{\label{thm.highdimth1}}
		Under the assumptions specified in Lemmas \ref{lem.highdimlemma1} and \ref{lem.highdimlemma2}, with regularization parameter $\lambda_n=2\xi\sqrt{\ln(d)/n}$, any of the solutions of equation \eqref{eq.HDPMTE} in the set $\mathbb{K}_{\boldsymbol{\beta}_0}=\{\boldsymbol{\beta}+\mathbf{\Delta}:\| \mathbf{\Delta} \|_2 \leq \sqrt{c_0R}/(12M\sqrt{s}) \}$, $\hat{\boldsymbol{\beta}}$, satisfies 
		\begin{align*}
		\|\hat{\boldsymbol{\beta}}-\boldsymbol{\beta}_0\|_2 \leq \frac{8\xi}{\kappa_1}\sqrt{\frac{s\ln(d)}{n}}
		\end{align*}
		with probability at least $1- \alpha_5 \exp(-\alpha_6 n \lambda^2_n)$ 
		for $n>\max\{64 \kappa_2^2 s \ln (d), \\
		96^2M^2 \xi^2 s^2\ln(d) 
		/(\kappa^2_1 c_0 R)\}$, where $\alpha_5$ and $\alpha_6$ are positive 
		constants.
	\end{theorem}
	The theorem implies that the convergence rate of $\hat{\boldsymbol{\beta}}$ depends on two critical quantities, the bound of the gradient of the loss function at the true parameter and the curvature of the loss function around the true parameter.  In particular, when the loss function becomes flatter at the true parameter and hence has a smaller bound of the gradient, the penalized MTE converges faster.  Similarly, when the loss function becomes more convex (i.e. larger curvature) in the restricted direction within the neighborhood of the true parameter (i.e., $\mathbb{C}(S)$), the penalized MTE also converges faster.
	
	However, as illustrated by Lemmas \ref{lem.highdimlemma1} and 
	\ref{lem.highdimlemma2}, the effects of $t$ on these two quantities are 
	often in the opposite directions.  For example, as $t$ increases, the 
	entire loss function generally becomes flatter which leads to a smaller 
	bound of the gradient at the true parameter.  But an increasing $t$ also 
	leads to a smaller curvature.  Therefore, selecting $t$ involves 
	controlling both the bound and the curvature.  To gain a faster convergence 
	rate, we need $t$ to be large to control the bound of the gradient, but 
	also need $t$ to be small to increase the curvature of the loss function.  
	Therefore, a trade-off has to be made when selecting $t$.  Note that when 
	$t>f(0)$, the penalized MTE becomes penalized minimum $\ell_2$ distance 
	estimation, therefore, we can see that the penalized MTE offers a more 
	refined trade-off between efficiency and robustness.  In Section 
	\ref{sec.algo}, we illustrate how to select $t$ in detail.

	%%%%%%%%%%%%%%%%%%%%%%%%%%%%%%%%%%%%%%%%%%%%%%%%%%%%%%%%%%%%%%%%%%%%%%%
	\section{Tuning Parameters and Algorithm}\label{sec.algo}

	%In this section, we discuss the implementation of the proposed penalized MTE, including the choices of tuning parameters, initial values, and computational algorithms. For the penalized MTE, there are two tuning parameters, $\lambda_{nj}$ that controls the regularization in the penalty function and $t$ that tunes the robustness in the loss function.  We could potentially treat them as a joint optimization problem that may be computationally intensive. Instead, in this article, we propose a simple method to select $\lambda_{nj}$ and a data-adaptive method to select $t$.
	
	\subsection{Choice of Regularization Parameter $\lambda$ and Connection to Robust Bayes}\label{subsec.tuninglambda}
	
	The performance of penalized estimator strongly relies on the choice of regularization parameter $\lambda$.  For fixed dimensional regression, in order to achieve oracle property, we adopt a simple BIC-type criterion \citep{Wang2007,Wang2013} to select $\lambda_{nj}$ that satisfies condition \eqref{eq.lambdacondt} by minimizing the following objective function
	\begin{align*}
	-\sum_{i=1}^{n}\ln_t\left(f({\bf z}_i;\boldsymbol{\beta})\right) +n 
	\sum_{j=1}^{d} \lambda_{nj} |\beta_j| - \ln(0.5n 
	\lambda_{nj})\ln(n), 
	\end{align*}
	which leads to the regularization parameter estimates
	\begin{align}
	\hat{\lambda}_{nj}=\frac{\ln(n)}{n |\tilde{\tilde{\beta}}_j|}, 
	\label{eq.optlambda}
	\end{align}
	where $\tilde{\tilde{\beta}}_j$ is an initial estimate of $\beta_j$. Note 
	that \eqref{eq.optlambda} can be viewed as a special case of adaptive-Lasso 
	penalty function. It is easy to see that this choice of $\lambda$ satisfies \eqref{eq.lambdacondt}, a necessary condition for the oracle property of the penalized MTE.
	
	In addition, since the traditional Lasso penalized estimation could be viewed as a Bayesian maximum a posteriori estimation (MAP) under independent Laplace (double-exponential) priors for $\beta_j$s.  Our penalized MTE could similarly be regarded as a new/robust version of MAP estimation where the traditional likelihood function is replaced by the tangent likelihood function.  Because of the robustness properties of the tangent likelihood, we expect to provide a robust posterior distribution. We leave this as a future research direction.
	
	For high-dimensional regression, we focus on the Lasso penalty function where the regularization parameter does not depend on $\beta_j$. Therefore we choose optimal $\lambda_n$ by minimizing median absolute prediction error through cross-validation over a grid.  
	
	\subsection{Choice of Tuning Parameter $t$}\label{subsec.tuningt}
	
\begin{figure}
		\centering
		\includegraphics[scale=0.4]{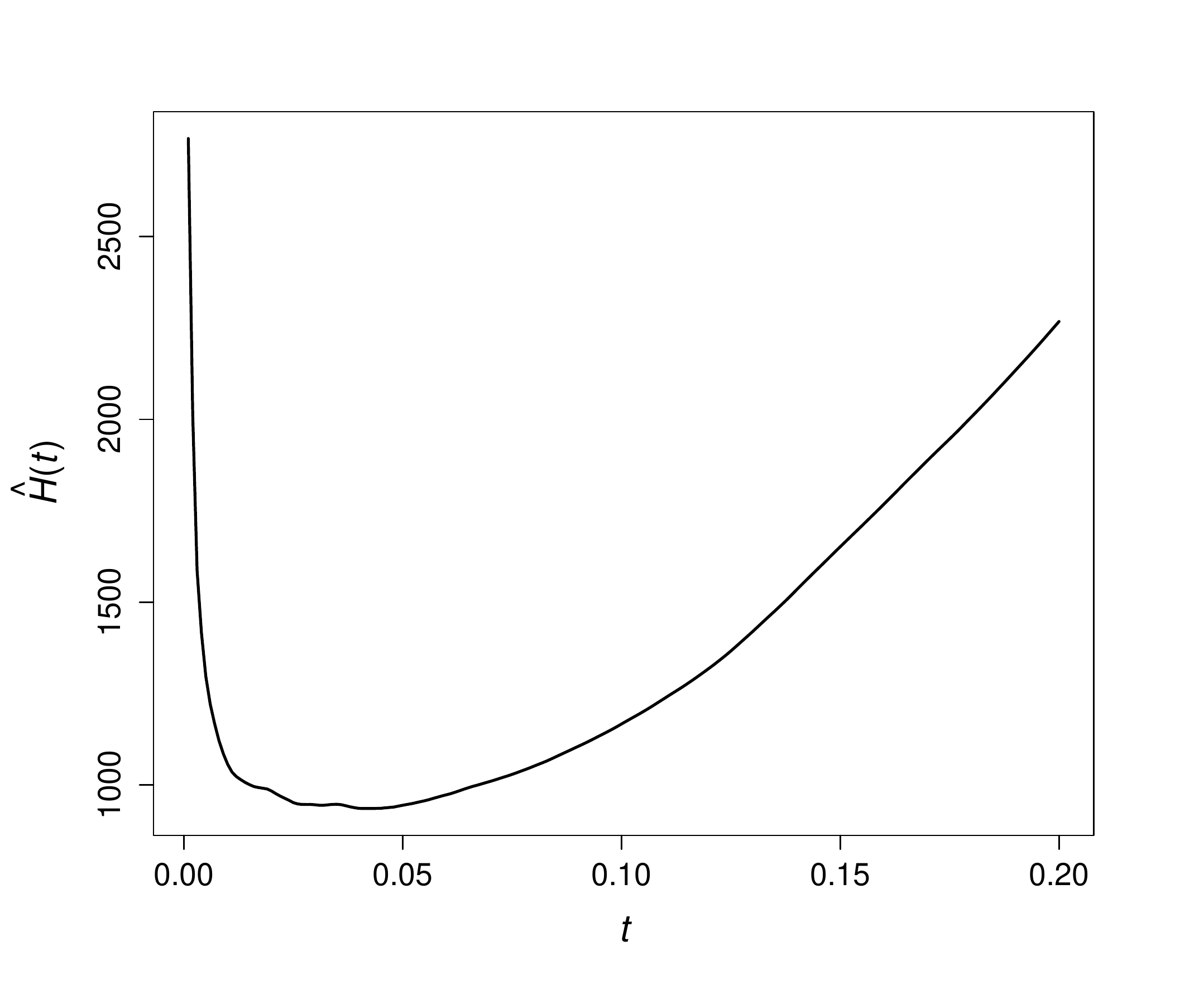}		
		\caption{Determinant of covariance matrix $\hat{H}(t)$ against $t$}	\label{fig.selct}	
	\end{figure}

	As discussed in Sections \ref{sec.MTEprop} and \ref{sec.asymPTE}, the tuning parameter $t$ controls the trade-off between robustness and efficiency, hence the choice of $t$ cannot be neglected. We use a simple data-driven method to grid search the optimal value of $t$ such that it minimizes the determinant of asymptotic covariance matrix of $\hat{\boldsymbol{\beta}}_S$ as in \eqref{eq.asycovmat}. 
	The idea of this approach is that $t$ is selected such that the proposed 
	estimator has minimum variance in order to achieve high efficiency. Similar 
	approach has been adopted by \citet{Wang2013} to select the tuning parameter 
	in the exponential squared loss function. As an illustration, Figure 
	\ref{fig.selct} shows one example of the value of the determinant of 
	\eqref{eq.asycovmat} denoted as $\hat{H}(t)$ against different values of 
	$t$. Note that under the high-dimensional regression setting, this grid 
	search method is applied when the number of nonzero $\beta$'s is less than $n$ in the iterative algorithm, which is often achieved after the first iteration.	

	\subsection{Choice of Initial Values}\label{subsec.init}
	
	When solving the optimization problem \eqref{eq.PMTE} and \eqref{eq.HDPMTE}, MTE could potentially lead to local maximums as the tangent likelihood loss function is nonconvex. Therefore, assigning suitable initial values for the optimization is critical. For our proposed method, we need to assign initial values for $\boldsymbol{\beta}$ as well as the preliminary scale estimate $\sigma^2_R$. For $\boldsymbol{\beta}$, we can use unpenalized LAD estimates as a candidate initial value because LAD is a monotone regression M-estimate whose objective function is always convex. For $\sigma^2_R$, we have adopted one of the well known robust scale parameter estimates, $\sigma_R=1.4826\times\textup{MAD}$, where MAD can be the median absolute deviance of residuals from LAD estimates, as the initial 	estimate. Other types of robust scale parameter estimation are also well developed and available \citep{rousseeuw1993} to serve as potential initial values.  

	\subsection{Computational Algorithm}\label{subsec.algm}
	
	Coordinate descent (CD) algorithm has recently been well recognized and appreciated for its surprisingly fast and efficient capability in solving $\ell_1$-regularization problem . It updates a single parameter one at a time while the rest are fixed. We choose the coordinate descent algorithm for its simplicity, speed and stability \citep{Wu2008, Friedman2007, Friedman2010,breheny2011}, and apply it for both fixed and high-dimensional regression settings.
	%Hence, it only requires $O(d)$ operations comparing to popular first-order method which requires $O(d^2)$ operations in updating the $d$-dimensional vector $\boldsymbol{\beta}$.
	We propose following 2-step iterative algorithm.
	
	\begin{enumerate}[label={\textbf{Step \arabic*}.}, leftmargin=*,start=1]
		{\bf \item Update tuning parameter $t$ and $\lambda_{nj}$}:		
		Given current estimates $\hat{\boldsymbol{\beta}}^{(j-1)}$, find optimal value $t^{(k)}$ such that $t^{(k)}$ minimizes the determinant of \eqref{eq.asycovmat} by grid search. Meanwhile, the optimal regularization parameter $\hat{\lambda}_{nj}^{(k)}$ can be calculated by \eqref{eq.optlambda}.
		
		{\bf \item Update parameter estimates}:		
		Based on $t^{(k)}$ and $\hat{\lambda}_{nj}^{(k)}$ that are obtained from Step 1, we use the coordinate descent algorithm to solve the optimization problem \eqref{eq.PMTE}.  Repeat Steps 1 and 2 until all elements of $\hat{\boldsymbol{\beta}}$ converge. Note that one may also update the scale parameter $\sigma_R^2$ based on the updated estimates of regression coefficient $\hat{\boldsymbol{\beta}}^{(k)}$ so that the estimation in next iteration is more accurate. 		
	\end{enumerate}
	
	This algorithm is directly applicable to both the fixed and high-dimensional regression settings with little modification (the optimal regularization parameter $\lambda_n$ is chosen by cross-validation, and need not to be updated between two steps). In practice, the range of $t$ in the grid-search procedure can be set from 0 to 0.2 in order to maintain high efficiency. From our limited numerical studies, the algorithm is computationally efficient with fast convergence. 
	%%%%%%%%%%%%%%%%%%%%%%%%%%%%%%%%%%%%%%%%%%%%%%%%%%%%%%%%%%%%%%%%%%%%%%%
	\section{Numerical Studies}\label{sec.numer} 
	
	\subsection{Monte Carlo Simulation for Fixed Dimensional Regressions}\label{subsubsec.simulow}
		
	For fixed dimensional regression, in order to achieve oracle estimates, we adopt the adaptive-Lasso penalty for MTE as well as its competitors, LAD \citep{Wang2007}, ESL \citep{Wang2013}, CQR \citep{Zou2008} and MLE \footnote{For CQR and MLE with adaptive-Lasso penalty, we directly employ the existing \texttt{R} packages \texttt{cqrReg} and \texttt{parcor}, respectively.} \citep{Zou2006}. The criteria used for comparison are median and median absolute deviation (MAD) of model error (ME) \citep{Fan2001} that is defined as
	\begin{align}
	\text{ME}=\frac{1}{n}(\hat{\boldsymbol{\beta}}-\boldsymbol{\beta}_0)^T\mathbf{X}^T\mathbf{X} (\hat{\boldsymbol{\beta}}-\boldsymbol{\beta}_0), \label{eq.modelerr}
	\end{align}
	and model selection errors which is measured by false negative rate (FNR) and false positive rate (FPR). Specifically, FNR is defined as the proportion of zero coefficient estimates whose corresponding true coefficients are nonzero, i.e., $\#\{j: \hat{\beta}_j=0, {\beta}_{0j} \neq 0\}/\#\{j: {\beta}_{0j} \neq 0\}$.  FPR is defined as the proportion of nonzero coefficient estimates whose corresponding true coefficients are zero, i.e., $\#\{j: \hat{\beta}_j \neq 0, {\beta}_{0j} = 0\}/\#\{j: {\beta}_{0j} = 0\}$.
	
	We set the true regression coefficient $\boldsymbol{\beta}_0 = (1,1.5,2,1,0,0,0,0, -2.5,-1,0,0)^T \in \mathbb{R}^{12}$, and consider following simulation designs: (1) $\epsilon_i \overset{\textup{iid}}{\sim} 0.7N(0,1)+0.3\textup{Unif}(-10,50)$ and $\mathbf{x}_i \overset{\textup{iid}}{\sim} N({\bf 0}, \mathbf{\Omega})$; 
	%(2) $\epsilon_i \overset{\textup{iid}}{\sim} \textup{Cauchy}$ and $\mathbf{x}_i \overset{\textup{iid}}{\sim} N({\bf 0}, \mathbf{\Omega})$; 
	(2) $\epsilon_i \overset{\textup{iid}}{\sim} 0.7N(0,1)+0.3N(10,10^2)$ and $\mathbf{x}_i \overset{\textup{iid}}{\sim} 0.8N({\bf 0}, \mathbf{I}) + 0.2N({\bf 3}, \mathbf{\Omega})$, where $\mathbf{I}$ is a $12\times 12$ identity matrix, and $\mathbf{\Omega}=\{ \Sigma_{ij} \}_{12 \times 12}$ is a $12\times 12$ covariance matrix with $\Sigma_{ij}=0.5^{|i-j|}$. Under each setting, we simulate 1000 Monte Carlo samples for different sample sizes, $n=100,200,400,800$. The results are reported in Tables \ref{table.simulow1} and \ref{table.simulow3}.

	\newcolumntype{L}[1]{>{\raggedright\arraybackslash}p{#1}}
	\newcolumntype{C}[1]{>{\centering\arraybackslash}m{#1}}
	\newcolumntype{R}[1]{>{\raggedleft\arraybackslash}p{#1}}
	\renewcommand{\arraystretch}{1.1}
	
	\begin{table}[h!]
		\centering
		\caption{Monte Carlo Simulation for regression models with error following mixture distribution: $\epsilon_i \overset{\text{iid}}{\sim} 0.7N(0,1)+0.3\textup{Unif}(-10,50)$ and covariates following distribution: $\mathbf{x}_i \overset{\text{iid}}{\sim} N({\bf 0}, \mathbf{\Omega})$.} \label{table.simulow1}	
		%\vspace{-0.4cm}
		\begin{tabular}{lccccr}
			\hline
			\hline 
			&&&&\multicolumn{2}{c}{Model Error} \\
			\cline{5-6}
			$n$ & Method  & FNR & FPR  & Median & MAD \\
			\hline
			100   
			& MTE		& 0.010		& 0.000		& 0.126		& 0.054 \\
			& LAD     	& 0.019		& 0.006		& 0.237		& 0.113 \\
			& ESL     	& 0.557		& 0.000		& 3.198		& 2.960 \\
			& CQR		& 0.343		& 0.234		& 10.951	& 1.367	\\
			& MLE		& 0.649		& 0.136		& 16.884	& 5.112 \\
			\hline
			200   
			& MTE		& 0.000		& 0.000		& 0.056		& 0.022 \\
			& LAD     	& 0.001		& 0.002		& 0.097		& 0.040 \\
			& ESL     	& 0.387		& 0.000		& 2.208		& 2.111 \\
			& CQR		& 0.334		& 0.204		& 10.202	& 0.923	\\
			& MLE		& 0.460		& 0.191		& 10.054	& 3.880 \\
			\hline
			400   
			& MTE		& 0.000		& 0.000		& 0.025		& 0.010 \\
			& LAD     	& 0.000		& 0.000		& 0.046		& 0.019 \\
			& ESL     	& 0.014		& 0.000		& 0.111		& 0.066 \\
			& CQR		& 0.333		& 0.169		& 9.932 	& 0.561	\\
			& MLE		& 0.286		& 0.220		& 4.877 	& 1.746 \\
			\hline
			800   
			& MTE		& 0.000		& 0.000		& 0.011		& 0.005 \\
			& LAD     	& 0.000		& 0.000		& 0.021		& 0.009 \\
			& ESL     	& 0.000		& 0.000		& 0.030		& 0.012 \\
			& CQR		& 0.333		& 0.141		& 9.818 	& 0.346	\\
			& MLE		& 0.175		& 0.225		& 2.627 	& 0.843 \\    
			\hline
			
		\end{tabular}
	\end{table}

	\begin{table}[h!]
		\centering
		\caption{Monte Carlo Simulation for regression models with random error following mixture distribution: $\epsilon_i \overset{\text{iid}}{\sim} 0.7N(0,1)+0.3N(10,10^2)$ and covariates following mixture distribution: $\mathbf{x}_i\overset{\text{iid}}{\sim} 0.8N({\bf 0}, \mathbf{I}) + 0.2N({\bf 3}, \mathbf{\Omega})$.} \label{table.simulow3}	
		%\vspace{-0.4cm}
		\begin{tabular}{lccccr}
			\hline
			\hline 
			&&&&\multicolumn{2}{c}{Model Error} \\
			\cline{5-6}
			$n$ & Method  & FNR & FPR  & Median & MAD \\
			\hline
			100   
			& MTE		& 0.009		& 0.001		& 0.126		& 0.058 \\
			& LAD     	& 0.011		& 0.007		& 0.263		& 0.122 \\
			& ESL     	& 0.654		& 0.000		& 9.336		& 9.073 \\
			& CQR		& 0.336		& 0.206		& 35.432	& 4.658	\\
			& MLE		& 0.306		& 0.255		& 7.584		& 2.743 \\
			\hline
			200   
			& MTE		& 0.000		& 0.000		& 0.057		& 0.023 \\
			& LAD     	& 0.000		& 0.002		& 0.125		& 0.051 \\
			& ESL     	& 0.278		& 0.000		& 2.269		& 2.144 \\
			& CQR		& 0.333		& 0.172		& 32.780	& 3.070	\\
			& MLE		& 0.137		& 0.295		& 4.639		& 1.372 \\
			\hline
			400   
			& MTE		& 0.000		& 0.000		& 0.025		& 0.010 \\
			& LAD     	& 0.000		& 0.001		& 0.066		& 0.026 \\
			& ESL     	& 0.000		& 0.000		& 0.085		& 0.033 \\
			& CQR		& 0.333		& 0.153		& 31.479	& 1.721	\\
			& MLE		& 0.051		& 0.294		& 3.042		& 0.781 \\
			\hline
			800   
			& MTE		& 0.000		& 0.000		& 0.012		& 0.005 \\
			& LAD     	& 0.000		& 0.001		& 0.043		& 0.015 \\
			& ESL     	& 0.000		& 0.000		& 0.027		& 0.017 \\
			& CQR		& 0.333		& 0.129		& 30.924	& 1.241	\\
			& MLE		& 0.008		& 0.267		& 2.257		& 0.467 \\     
			\hline
			
		\end{tabular}
	\end{table}

	As Tables \ref{table.simulow1} and \ref{table.simulow3} illustrate,  MTE outperforms all other methods in terms of model errors and variable selection accuracy.  As the sample size $n$ increases, the performance of all methods improve, but MTE dominates all other methods uniformly.  
	
	\subsection{Monte Carlo Simulation for High Dimensional Regressions}\label{subsubsec.simuhigh}
	
	We further demonstrate the performance of MTE under high-dimensional regression settings with $d=500$ through a Monte Carlo simulation.  We set the true coefficient $\boldsymbol{\beta}_0=(3,1.5,2,-2.5,-2,3,1.5,2,-2.5,-2,\dotso,0)^T \in \mathbb{R}^{500}$, a 500-dimensional coefficient vector with 3 non-zeros. We conduct 100 Monte Carlo simulations from model \eqref{eq.lreg} with sample size $n=200$.  We consider three types of covariates: (1) ${\bf x}_i \overset{\textup{iid}}{\sim}N({\bf 0},\mathbf{I})$; (2) ${\bf x}_i \overset{\textup{iid}}{\sim} N({\bf 0}, \mathbf{\Omega})$; and (3) ${\bf x}_i \overset{\textup{iid}}{\sim} 0.8N({\bf 0},\mathbf{I}) + 0.2N({\bf 3}, \mathbf{\Omega})$, where $\mathbf{I}$ is a $d\times d$ identity matrix, and $\mathbf{\Omega}=\{\Sigma_{ij}\}_{d \times d}$ with $\Sigma_{ij}=0.5^{|i-j|}$. We also consider six types of random errors:
	\begin{enumerate}
		\setlength\itemsep{-0.5em}
		\item[(1)] $\epsilon_i \overset{\textup{iid}}{\sim} N(0,1)$;
		\item[(2)] $\epsilon_i \overset{\textup{iid}}{\sim} 0.8N(0,1)+0.2N(0, 20^2)$;
		\item[(3)] $\epsilon_i \overset{\textup{iid}}{\sim} 0.8N(0,1)+0.2N(50, 10^2)$;
		\item[(4)] $\epsilon_i \overset{\textup{iid}}{\sim} 0.6N(0,1)+0.2N(20, 10^2)+0.2N(-50, 10^2)$;
		\item[(5)] $\epsilon_i \overset{\textup{iid}}{\sim} \text{Cauchy}$;
		\item[(6)] $\epsilon_i \overset{\textup{iid}}{\sim} t(2)$.
	\end{enumerate}
	
\begin{figure}
		\centering 		
		\vspace{-0.4cm}
		\includegraphics[scale=0.8]{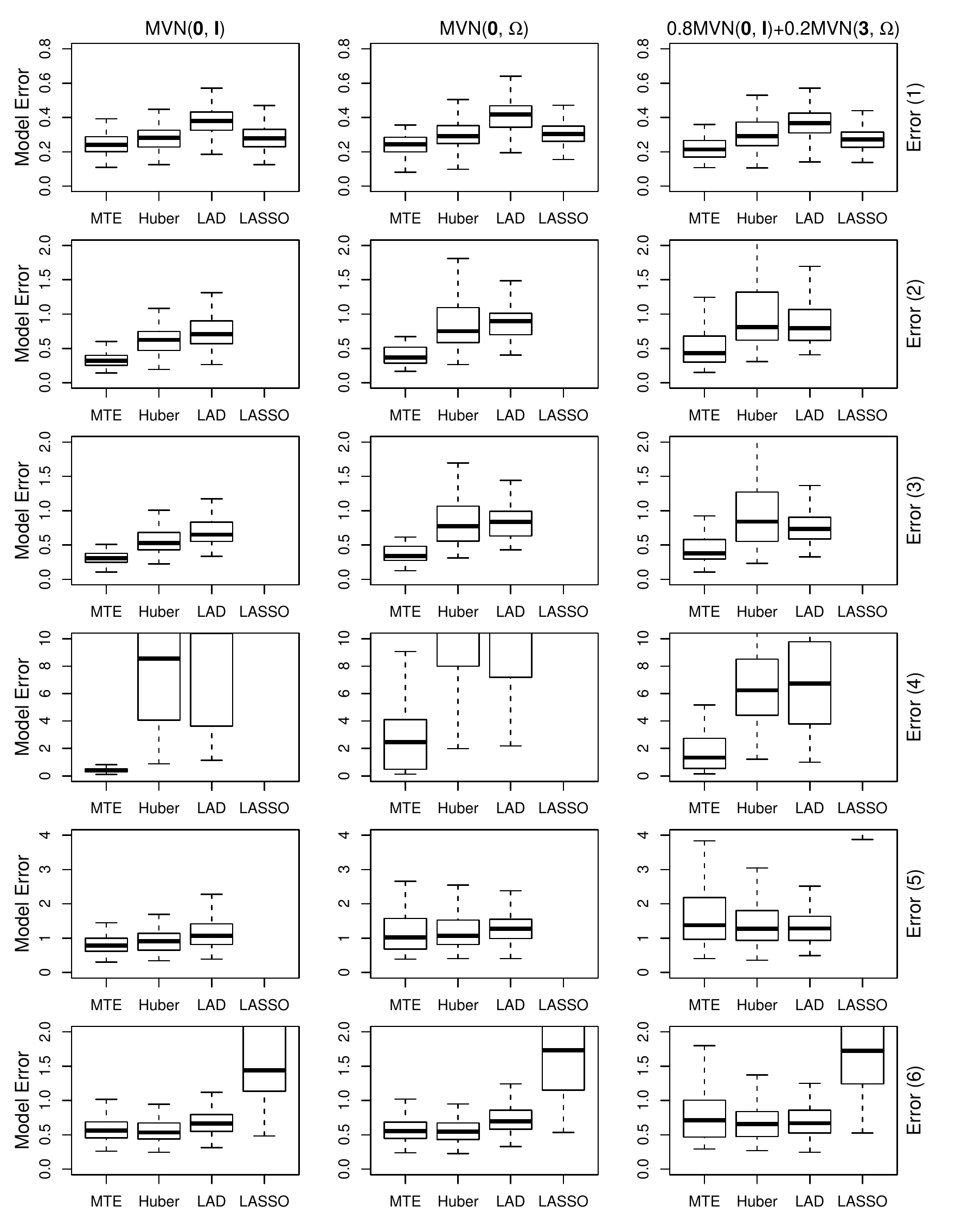}
		\caption{Box plots of model errors for different methods. Six types of errors are in row direction and three types of covariates are in column direction.} \label{fig.highdim1}
	\end{figure}

	 We compare our methods to famous robust estimators, Huber \citep{fan2016huber} and LAD \citep{WangLie2013}. All methods are equipped with Lasso penalty function. We also add traditional LASSO (implemented using \texttt{R} package \texttt{parcor}) in the comparison. The optimal tuning parameter $\lambda$ is chosen by minimizing median absolute prediction error through cross-validation. Figure \ref{fig.highdim1} shows the box plots of model errors. The range of vertical axis is truncated from above for better comparison. As we can see, traditional LASSO estimator fails when the data is contaminated. For the rest three robust estimator, MTE performs the best in most scenarios. We exclude CQR in the comparison because the \texttt{R} package \texttt{cqrReg} yields poor performance using the default algorithm and may not be appropriate for high-dimensional settings. We do not include ESL because to our best knowledge, there is no published work that studies ESL in high-dimensional regression.
	
	We also report mean, median and MAD of model errors in Table \ref{table.simuhigh1}. In addition, we further investigate the variable selection accuracy, and report the averaged counts of true positive covariates (TP) and false positive covariates (FP), i.e., $\textup{TP}=\#\{j: \hat{\beta}_j \neq 0, {\beta}_{0j} \neq 0\}$ and $\textup{FP}=\#\{j: \hat{\beta}_j \neq 0, {\beta}_{0j} = 0\}$.   
	
	\begin{table}
		\centering
		\caption{Comparison of MTE, Huber, LAD and LASSO on model error and variable selection accuracy under high-dimensional regression setting with $n=200, d=500$. TP is the average count of correctly estimated nonzero coefficients; and FP is the average count of nonzero estimates whose corresponding true coefficients are zero.  Note that there are 10 nonzero and 490 zero true coefficients in total. The average is based on 100 Monte Carlo simulations.} \label{table.simuhigh1}	
		%\vspace{-0.4cm}
		%\resizebox{\textwidth}{!}{%
		%\begin{tabular}{|L{0.6cm} | L{1.1cm} | C{0.8cm} C{0.8cm} C{0.8cm} C{0.6cm} C{0.6cm} | C{0.8cm} C{0.8cm} C{0.8cm} C{0.6cm} C{0.6cm} | C{0.8cm} C{0.8cm} C{0.8cm} C{0.6cm} C{0.6cm}|}
		\resizebox{\textwidth}{!}{%
		\begin{tabular}{|l|l|ccccc|ccccc|ccccc|}
			\hline			
			$\epsilon$ &  &\multicolumn{5}{c|}{${\bf x}_i \overset{\textup{iid}}{\sim}N({\bf 0}, \mathbf{I})$} & \multicolumn{5}{c|}{${\bf x}_i \overset{\textup{iid}}{\sim} N({\bf 0}, \mathbf{\Omega})$} & \multicolumn{5}{c|}{${\bf x}_i \overset{\textup{iid}}{\sim} 0.8N({\bf 0},\mathbf{I}) + 0.2N({\bf 3}, \mathbf{\Omega})$}\\
				\cline{3-17}
				& & Mean & Med. & MAD & TP & FP & Mean & Med. & MAD & TP & FP & Mean & Med. & MAD & TP & FP \\
				\hline
				$\epsilon(1)$ & MTE & 0.24 & 0.24 & 0.04 & 10.0 & 28.4 & 0.29 & 0.24 & 0.04 & 9.9 & 26.9 & 0.25 & 0.21 & 0.05 & 10.0 & 18.6 \\ 
				& Huber & 0.28 & 0.28 & 0.05 & 10.0 & 29.5 & 0.31 & 0.29 & 0.05 & 10.0 & 29.9 & 0.33 & 0.29 & 0.06 & 9.9 & 26.3 \\ 
				& LAD & 0.37 & 0.38 & 0.05 & 10.0 & 48.2 & 0.41 & 0.42 & 0.06 & 10.0 & 55.0 & 0.37 & 0.37 & 0.06 & 10.0 & 51.5 \\ 
				& Lasso & 0.28 & 0.28 & 0.05 & 10.0 & 41.6 & 0.30 & 0.30 & 0.04 & 10.0 & 43.9 & 0.28 & 0.27 & 0.04 & 10.0 & 43.8 \\
				\hline
				$\epsilon(2)$ & MTE & 0.33 & 0.32 & 0.07 & 10.0 & 21.9 & 0.68 & 0.37 & 0.10 & 9.9 & 26.3 & 0.64 & 0.43 & 0.14 & 9.9 & 20.7 \\ 
				& Huber & 0.64 & 0.62 & 0.14 & 10.0 & 25.4 & 1.12 & 0.75 & 0.22 & 9.9 & 28.2 & 1.05 & 0.81 & 0.26 & 9.9 & 25.3 \\ 
				& LAD & 0.77 & 0.71 & 0.16 & 10.0 & 47.2 & 0.93 & 0.89 & 0.16 & 10.0 & 50.4 & 0.88 & 0.79 & 0.20 & 10.0 & 48.0 \\ 
				& Lasso & 21.20 & 19.97 & 4.63 & 8.3 & 33.5 & 21.45 & 20.96 & 3.56 & 6.3 & 27.1 & 16.49 & 16.36 & 1.91 & 3.9 & 19.1 \\ 
				\hline
				$\epsilon(3)$ & MTE & 0.31 & 0.30 & 0.06 & 10.0 & 23.4 & 0.78 & 0.34 & 0.08 & 9.8 & 31.8 & 0.58 & 0.38 & 0.12 & 9.8 & 24.6 \\ 
				& Huber & 0.57 & 0.53 & 0.11 & 10.0 & 26.6 & 1.16 & 0.77 & 0.22 & 9.9 & 33.4 & 1.00 & 0.84 & 0.35 & 9.8 & 35.0 \\ 
				& LAD & 0.71 & 0.65 & 0.13 & 10.0 & 51.4 & 0.83 & 0.84 & 0.18 & 10.0 & 58.4 & 0.76 & 0.73 & 0.15 & 10.0 & 56.3 \\ 
				& Lasso & 48.21 & 48.55 & 3.57 & 0.4 & 1.0 & 45.78 & 46.79 & 2.61 & 0.5 & 1.2 & 24.89 & 24.62 & 3.88 & 0.4 & 2.8 \\ 
				\hline
				$\epsilon(4)$ & MTE & 1.01 & 0.39 & 0.13 & 9.8 & 16.5 & 2.91 & 2.46 & 1.95 & 9.2 & 22.3 & 1.78 & 1.34 & 0.88 & 9.4 & 29.1 \\ 
				& Huber & 11.12 & 8.54 & 5.50 & 9.0 & 23.5 & 13.19 & 12.42 & 4.50 & 7.8 & 22.7 & 6.51 & 6.24 & 1.89 & 8.2 & 29.5 \\ 
				& LAD & 12.34 & 10.46 & 7.27 & 8.8 & 37.4 & 12.88 & 11.66 & 5.64 & 8.2 & 38.4 & 7.08 & 6.72 & 2.93 & 8.3 & 35.7 \\ 
				& Lasso & 50.70 & 50.16 & 4.26 & 0.7 & 3.9 & 47.81 & 47.50 & 3.77 & 0.6 & 3.7 & 27.35 & 27.24 & 4.20 & 0.3 & 5.0 \\ 
				\hline
				$\epsilon(5)$ & MTE & 0.86 & 0.79 & 0.19 & 10.0 & 22.4 & 1.38 & 1.02 & 0.42 & 9.8 & 25.0 & 1.66 & 1.38 & 0.57 & 9.8 & 34.1 \\ 
				& Huber & 0.97 & 0.91 & 0.25 & 10.0 & 28.2 & 1.29 & 1.07 & 0.28 & 9.9 & 30.1 & 1.42 & 1.28 & 0.39 & 9.8 & 31.7 \\ 
				& LAD & 1.15 & 1.07 & 0.27 & 10.0 & 47.1 & 1.37 & 1.28 & 0.28 & 10.0 & 52.2 & 1.32 & 1.28 & 0.35 & 10.0 & 46.7 \\ 
				& Lasso & 35.90 & 40.87 & 12.12 & 4.0 & 14.1 & 35.00 & 40.36 & 10.78 & 3.1 & 13.1 & 21.09 & 20.67 & 6.97 & 2.2 & 9.6 \\
				\hline
				$\epsilon(6)$ & MTE & 0.59 & 0.56 & 0.12 & 10.0 & 26.7 & 0.71 & 0.55 & 0.12 & 9.9 & 26.8 & 0.88 & 0.71 & 0.26 & 9.9 & 23.7 \\ 
				& Huber & 0.56 & 0.53 & 0.11 & 10.0 & 29.2 & 0.60 & 0.55 & 0.12 & 10.0 & 27.5 & 0.72 & 0.65 & 0.18 & 9.9 & 29.0 \\ 
				& LAD & 0.69 & 0.66 & 0.12 & 10.0 & 50.1 & 0.72 & 0.69 & 0.14 & 10.0 & 52.3 & 0.70 & 0.67 & 0.15 & 10.0 & 50.0 \\ 
				& Lasso & 2.96 & 1.44 & 0.44 & 9.9 & 38.7 & 3.42 & 1.73 & 0.71 & 9.8 & 43.1 & 2.64 & 1.72 & 0.62 & 9.7 & 41.8 \\
				\hline
				
			\end{tabular}
		}
		\end{table}

		\subsection{Real Data Examples}\label{subsec.realdataeg}
		
		We demonstrate the performance of the proposed penalized MTE using some real data examples. We first apply it to Boston housing price dataset (\url{https://archive.ics.uci.edu/ml/datasets/Housing}), which is commonly used as an example for regressions. It is particularly of interest for robust regression analysis as the dataset contains outliers and skewed variables. There are 14 variables in total: \textit{medv}, \textit{rm}, \textit{tax}, \textit{ptratio}, \textit{lstat}, \textit{nox}, \textit{dis}, \textit{crim}, \textit{zn}, \textit{indus}, \textit{age}, \textit{black}, \textit{chas}, \textit{rad}.  Detailed explanations of these variables can be found in the supplementary materials.  We use \textit{medv} (median house price) as the response variable.  Following \citet{Wu2010} and references therein, we take logarithm of variables \textit{crim}, \textit{lstat} and \textit{tax}, and standardize all variables before fitting the model. Table \ref{table.realdata1} gives the variable selection results. Standard errors are obtained based on 500 bootstrapping samples. We find that the traditional adaptive-Lasso (MLE) selects many (10 out of 13) variables. MTE and CQR select 5 variables \textit{rm}, \textit{ln(tax)}, \textit{ptratio}, \textit{ln(stat)}, and \textit{dis}. This finding is largely consistent with variables commonly used in the literature. For example, four variables \textit{rm}, \textit{ln(tax)}, \textit{ptratio}, and \textit{ln(stat)} are considered in \citet{Opsomer1998}, \citet{YuLu2004} and \citet{Wu2010}, whereas three variables \textit{rm}, \textit{ln(stat)}, \textit{dis} are used in \citet{Chaudhuri1997}.       		
				
		%\begin{center}
		%\renewcommand{\arraystretch}{1.1}
		%\begin{spacing}{1}
		\begin{table}[h!]
			\centering
			\caption{Coefficients estimates of Boston housing price data using different methods. The standard errors of coefficient estimates are in parenthesis and they are based on 500 bootstrap samples. ``0'' indicates that the corresponding variable is not selected.} \label{table.realdata1}
			%\vspace{-0.4cm}
			\resizebox{\textwidth}{!}{%
			\begin{tabular}{lccccc}	
				\hline 
				\hline
				Variable 	& MTE  	& LAD   & ESL   & CQR   & MLE  \\
				\hline
				rm        & 0.379 (\textit{0.108})  & 0.323 (\textit{0.134})  & 0.308 (\textit{0.209})   & 0.448 (\textit{0.146})  & 0.200 (\textit{0.063})\\
				ln(tax)    & -0.131 (\textit{0.070}) & 0       & 0       & -0.019 (\textit{0.034}) & -0.134 (\textit{0.044})\\
				ptratio   & -0.161 (\textit{0.031}) & -0.156 (\textit{0.060})  & -0.130 (\textit{0.071})      & -0.083 (\textit{0.036})  & -0.201 (\textit{0.026})\\
				ln(lstat)  & -0.436 (\textit{0.078}) & -0.436 (\textit{0.125})  & -0.453 (\textit{0.177})  & -0.453 (\textit{0.119})  & -0.609 (\textit{0.077})\\
				nox       & 0 	   & 0		 & 0	   & 0		 & -0.152 (\textit{0.045})\\
				dis       & -0.069 (\textit{0.068}) & 0 		 & 0	   & -0.025 (\textit{0.038})  & -0.233 (\textit{0.043})\\
				ln(crim)   & 0 	   & 0		 & 0	   & 0		 & 0\\
				zn        & 0 	   & 0		 & 0	   & 0		 & 0\\
				indus     & 0 	   & 0		 & 0	   & 0		 & 0\\
				age       & 0 	   & 0		 & 0	   & 0		 & 0.037 (\textit{0.052})\\
				black     & 0 	   & 0		 & 0	   & 0       & 0.078 (\textit{0.029})\\
				chas      & 0 	   & 0		 & 0	   & 0       & 0.054 (\textit{0.036})\\
				rad       & 0 	   & 0		 & 0	   & 0  	 & 0.140 (\textit{0.060})\\
				\hline
				
			\end{tabular}
		}
		\end{table}	
		%\end{spacing}
		
		%\end{center}

		Next, we apply the proposed method to an expression quantitative trait loci (eQTL) dataset under a high-dimensional regression. The dataset can be accessed at NCBI Gene Expression Omnibus data repository (\url{http://www.ncbi.nlm.nih.gov/geo}) with access number GSE3330.  The dataset contains a sample of $n=60$ individuals of  F2-ob/ob(B) mice with 22,575 different Affymetrix probe sets. The expression value for each prob set is microarray-derived gene expression measurements (mRNA abundance traits), and they are obtained using the Affymetrix MOE430B microarrays (Array B of GeneChip Mouse Expression Set 430).  \citet{lan2006} developed and studied this sample to identify regulatory networks. We investigate the linear relationship of gene expressions and PEPCK, the numbers of phosphoenopyruvate carboxykinase (NM\_011044) measured by quatitative real-time RT-PCR. Similar study has been done by \citep{song2015}.  First, we pre-screened all 22,575 probes variables by calculating the correlation coefficients with the response variable PEPCK.  We use 1000 gene expression variables who have the highest marginal correlation to repsonse variable as covariates. We compare our method with some alternatives, LAD-Lasso, Huber-Lasso, and LASSO.
		
		%In order to have a fair comparison, the regularization parameters for 
		%all methods are chosen such that the maximum number of selected probe 
		%sets is 5. 
		%Table \ref{table.realdatahigh1} shows the probe selection results and coefficient estimates for different methods. 
		MTE selects four probe sets: ``1438937\_x\_at", ``1437871\_at", ``1439163\_at", and ``1439617\_s\_at". 
		Among them, ``1438937\_x\_at" is the common one that has been selected by all methods, and ``1437871\_at" has been selected by three methods. More importantly, the four selected probe sets by MTE are all covered by LASSO, which has selected five probe sets. The selection results from LAD and Huber, however, are very different from MTE and LASSO. By exploratory analysis, we found that the response variable in this dataset is little contaminated. In this case, as we expected, MTE and LASSO should produce similar estimates.
		%The selected probes from LAD and Huber estimators are less consistent with LASSO as well as the coefficient estimates. 
		
		%\begin{table}[h]
		%	\centering
		%	\caption{Variable selection results of the eQTL Gene Expression data. Four methods are used for comparison} \label{table.realdatahigh1}
			%\vspace{-0.4cm}
		%	\begin{tabular}{|l|l|}
		%		\hline
		%		Methods & Selected Genes\\
		%		\hline
		%		MTE     & ``1438937\_x\_at", ``1437871\_at", ``1439163\_at", ``1439617\_s\_at" \\
		%		\hline
		%		LAD		& ``1438937\_x\_at", ``1437040\_at", ``1431742\_at", ``1437085\_at", ``1430239\_at" \\
		%		\hline
		%		Huber		& ``1438937\_x\_at", ``1437871\_at", ``1430239\_at"\\
		%		\hline
		%		LASSO	& ``1438937\_x\_at", ``1437871\_at", ``1437040\_at", ``1439163\_at" ``1439617\_s\_at"\\
		%		\hline
		%		
		%	\end{tabular}
		%\end{table}
		
%		\begin{table}[h]
%			\centering
%			\caption{Variable selection results of the eQTL Gene Expression data. Four methods are used for comparison} \label{table.realdatahigh1}
%			\begin{tabular}{|l|c|c|c|c|}
%				\hline
%				Probe ID & MTE & LAD & Huber & LASSO\\
%				\hline
%				1438937\_x\_at & 0.145 & 0.074 & 0.150 & 0.138\\
%				1437871\_at & -0.075 & & -0.137 & -0.070\\
%				1439163\_at & 0.053 & & & 0.049\\
%				1439617\_s\_at & 0.025 &  & & 0.019\\
%				1437040\_at  & & 0.013 & & 0.023\\
%				1431742\_at  & & 0.012 & & \\
%				1437085\_at  & & -0.004 & & \\
%				1430239\_at  & & 0.018 & 0.039 & \\
%				1428692\_at  & & & 0.002 & \\
%				\hline
%			\end{tabular}
%		\end{table}
		
		We further evaluate the out-of-sample prediction performance of these methods. The dataset is randomly split to training set (54 observations) and testing set (6 observations). Table \ref{table.realdatahigh2} reports the average mean squared prediction error (MSPE) and average model size, i.e. number of significant genes, over 100 random splits. From Table \ref{table.realdatahigh2}, we can see that the out-of-sample prediction performance of MTE is uniformly better than the other methods. We notice that the standard deviation of model size (number of selected variables) of MTE is also the smallest among all methods.
		
		\begin{table}
			\centering
			\caption{Mean squared prediction errors (MSPE) and model sizes obtained from different methods using the eQTL dataset. The average MSPE and model size based on 100 random splits are reported. Numbers in the parenthesis are standard errors.} \label{table.realdatahigh2}
			%\vspace{-0.4cm}
			\begin{tabular}{lcc}
				\hline
				\hline
				Methods & MSPE 					 & Model Size\\
				\hline
				MTE     & 0.565 (\textit{0.034}) & 5.58 (\textit{1.210})\\
				LAD		& 0.683 (\textit{0.038}) & 5.02 (\textit{1.461})\\
				Huber   & 0.574 (\textit{0.034}) & 6.16 (\textit{1.436})\\
				LASSO	& 0.712 (\textit{0.039}) & 5.80 (\textit{3.296})\\
				\hline
				
			\end{tabular}
		\end{table}

		%%%%%%%%%%%%%%%%%%%%%%%%%%%%%%%%%%%%%%%%%%%%%%%%%%%%%%%%%%%%%%%%%%%%%%
		\section{Conclusion}\label{sec.concl}
		
		We have proposed a new class of robust mean regression estimators that can produce robust and efficient estimates. Our proposed maximum tangent likelihood estimate (MTE) covers a number of existing estimators, such as MLE, minimum distance estimator, Mallows type estimator, and trimmed likelihood estimator as special cases. More interestingly, we show that solving the proposed MTE is equivalent to minimizing a combination of Kullback-Leibler (KL) and $\ell_2$ distance, where the weights depend on the choice of tuning parameter $t$. Our proposed penalized maximum tangent likelihood estimator performs well in robust estimation and variable selection under both fixed and high-dimensional regression. In addition to various numerical studies that demonstrate superior performance in practice, we have shown that the unpenalized MTE enjoys nice theoretical properties such as consistency and asymptotic normality, and the oracle property holds for the penalized MTE under fixed dimensional regression. Further, we show that under an ultra-high-dimensional regression setting when $d$ can grow exponentially with $n$, for any positive $t$, the penalized MTE is consistent in the optimal order of $\sqrt{\ln(d)/n}$.

	%%%%%%%%%%%%%%%%%%%%%%%%%%%%%%%%%%%%%%%%%%%%%%%%%%%%%%%%%%%%%%%%%%%%%%

	\bibliographystyle{apalike}
	\bibliography{YichenBib}
\end{document}